\documentclass{aa}  
\usepackage{graphicx}
\usepackage{txfonts}
\usepackage{hyperref}
\hypersetup{               
    colorlinks=true,
    linkcolor= blue,
    citecolor=blue
}
\RequirePackage[normalem]{ulem} 
\RequirePackage{color}\definecolor{RED}{rgb}{1,0,0}\definecolor{BLUE}{rgb}{0,0,1} 
\providecommand{\DIFaddbegin}{} 
\providecommand{\DIFaddend}{} 
\providecommand{\DIFdelbegin}{} 
\providecommand{\DIFdelend}{} 
\providecommand{\DIFaddbeginFL}{} 
\providecommand{\DIFaddendFL}{} 
\providecommand{\DIFdelbeginFL}{} 
\providecommand{\DIFdelendFL}{} 
\newcommand{\DIFscaledelfig}{0.5}
\RequirePackage{settobox} 
\RequirePackage{letltxmacro} 
\newsavebox{\DIFdelgraphicsbox} 
\newlength{\DIFdelgraphicswidth} 
\newlength{\DIFdelgraphicsheight} 
\LetLtxMacro{\DIFOincludegraphics}{\includegraphics} 
\newcommand{\DIFaddincludegraphics}[2][]{{\color{blue}\fbox{\DIFOincludegraphics[#1]{#2}}}} 
\newcommand{\DIFdelincludegraphics}[2][]{
        \sbox{\DIFdelgraphicsbox}{\DIFOincludegraphics[#1]{#2}}
        \settoboxwidth{\DIFdelgraphicswidth}{\DIFdelgraphicsbox} 
        \settoboxtotalheight{\DIFdelgraphicsheight}{\DIFdelgraphicsbox} 
        \scalebox{\DIFscaledelfig}{
                \parbox[b]{\DIFdelgraphicswidth}{\usebox{\DIFdelgraphicsbox}\\[-\baselineskip] \rule{\DIFdelgraphicswidth}{0em}}\llap{\resizebox{\DIFdelgraphicswidth}{\DIFdelgraphicsheight}{
                                \setlength{\unitlength}{\DIFdelgraphicswidth}
                                \begin{picture}(1,1)
                                \thicklines\linethickness{2pt} 
                                {\color[rgb]{1,0,0}\put(0,0){\framebox(1,1){}}}
                                {\color[rgb]{1,0,0}\put(0,0){\line( 1,1){1}}}
                                {\color[rgb]{1,0,0}\put(0,1){\line(1,-1){1}}}
                                \end{picture}
                        }\hspace*{3pt}}} 
} 
\LetLtxMacro{\DIFOaddbegin}{\DIFaddbegin} 
\LetLtxMacro{\DIFOaddend}{\DIFaddend} 
\LetLtxMacro{\DIFOdelbegin}{\DIFdelbegin} 
\LetLtxMacro{\DIFOdelend}{\DIFdelend} 
\DeclareRobustCommand{\DIFaddbegin}{\DIFOaddbegin \let\includegraphics\DIFaddincludegraphics} 
\DeclareRobustCommand{\DIFaddend}{\DIFOaddend \let\includegraphics\DIFOincludegraphics} 
\DeclareRobustCommand{\DIFdelbegin}{\DIFOdelbegin \let\includegraphics\DIFdelincludegraphics} 
\DeclareRobustCommand{\DIFdelend}{\DIFOaddend \let\includegraphics\DIFOincludegraphics} 
\LetLtxMacro{\DIFOaddbeginFL}{\DIFaddbeginFL} 
\LetLtxMacro{\DIFOaddendFL}{\DIFaddendFL} 
\LetLtxMacro{\DIFOdelbeginFL}{\DIFdelbeginFL} 
\LetLtxMacro{\DIFOdelendFL}{\DIFdelendFL} 
\DeclareRobustCommand{\DIFaddbeginFL}{\DIFOaddbeginFL \let\includegraphics\DIFaddincludegraphics} 
\DeclareRobustCommand{\DIFaddendFL}{\DIFOaddendFL \let\includegraphics\DIFOincludegraphics} 
\DeclareRobustCommand{\DIFdelbeginFL}{\DIFOdelbeginFL \let\includegraphics\DIFdelincludegraphics} 
\DeclareRobustCommand{\DIFdelendFL}{\DIFOaddendFL \let\includegraphics\DIFOincludegraphics} 

\begin{document} 

   \title{Identifying the energy release site in a Solar microflare with a jet}

   \author{
          Andrea Francesco Battaglia \inst{1, 2} 
          \and 
          Wen Wang \inst{1, 3}
          \and
          Jonas Saqri \inst{4}
          \and
          Tatiana Podladchikova \inst{5}
          \and
          Astrid M. Veronig \inst{4, 6}
          \and
          Hannah Collier \inst{1, 2}
          \and
          Ewan C. M. Dickson \inst{4}
          \and
          Olena Podladchikova \inst{7}
          \and
          Christian Monstein \inst{8}
          \and
          Alexander Warmuth \inst{7}
          \and
          Frédéric Schuller \inst{7}
          \and
          Louise Harra \inst{2, 9}
          \and
          Säm Krucker \inst{1, 10}
          }

   \institute{
    Institute for Data Science, University of Applied Sciences and Arts Northwestern Switzerland (FHNW), Bahnhofstrasse 6, 5210 Windisch, Switzerland \\
    \email{andrea.battaglia@fhnw.ch}
    \and
    Institute for Particle Physics and Astrophysics (IPA), Swiss Federal Institute of Technology in Zurich (ETHZ), Wolfgang-Pauli-Strasse 27, 8039 Zurich, Switzerland 
    \and
    School of Earth and Space Sciences, Peking University, Beijing, 100871, China
    \and
    Institute of Physics, University of Graz, Universit\"atsplatz 5, A-8010 Graz, Austria
    \and
    Skolkovo Institute of Science and Technology, Bolshoy Boulevard 30, bld. 1, 121205, Moscow, Russia
    \and
    Kanzelh\"ohe Observatory for Solar and Environmental Research, University of Graz, Kanzelh\"ohe 19, 9521 Treffen, Austria
    \and
    Leibniz-Institut f\"ur Astrophysik Potsdam (AIP), An der Sternwarte 16, D-14482 Potsdam, Germany
    \and
    IRSOL Istituto Ricerche Solari "Aldo e Cele Daccò", Università della Svizzera italiana, via Patocchi 57, 6605 Locarno, Switzerland
    \and
    Physikalisch-Meteorologisches Observatorium Davos, World Radiation Center, 7260 Davos Dorf, Switzerland
    \and
    Space Sciences Laboratory, University of California, 7 Gauss Way, 94720 Berkeley, USA
             }
\authorrunning{A. F. Battaglia et~al.}

   \date{Received September 19, 2022; accepted December 12, 2022}

 
  \abstract
   {One of the main science questions of the Solar Orbiter and Parker Solar Probe missions deals with understanding how electrons in the lower solar corona are accelerated and how they subsequently access interplanetary space.}
   {We aim to investigate the electron acceleration and energy release sites as well as the manner in which accelerated electrons access the interplanetary space in the case of the SOL2021-02-18T18:05 event, a GOES A8 class microflare associated with a coronal jet.}
   {This study takes advantage of three different vantage points, Solar Orbiter, STEREO-A, and Earth, with observations drawn from eight different instruments, ranging from radio to X-ray. Multi-wavelength timing analysis combined with UV/EUV imagery and X-ray spectroscopy by Solar Orbiter/STIX (Spectrometer/Telescope for Imaging X-rays) is used to investigate the origin of the observed emission during different flare phases.}
   {The event under investigation satisfies the classical picture of the onset time of the acceleration of electrons coinciding with the jet and the radio type III bursts. This microflare features prominent hard X-ray (HXR) nonthermal emission down to at least 10 keV and a spectrum that is much harder than usual
for a microflare with $\gamma = 2.9 \pm  0.3$.
   From Earth's vantage point, the microflare is seen near the limb, revealing the coronal energy release site above the flare loop in EUV, which, from STIX spectroscopic analysis, turns out to be hot (i.e., at roughly the same temperature of the flare). Moreover, this region is moving toward higher altitudes over time ($\sim 30\,\textrm{km}\,\textrm{s}^{-1}$). During the flare, the same region spatially coincides with the origin of the coronal jet.
   Three-dimensional (3D) stereoscopic reconstructions of the propagating jet highlight that the ejected plasma moves along a curved trajectory.
   }
   {Within the framework of the interchange reconnection model, we conclude that the energy release site observed above-the-loop corresponds to the electron acceleration site, corroborating that interchange reconnection is a viable candidate for particle acceleration in the low corona on field lines open to interplanetary space.
   }


   \maketitle
%

\section{Introduction}

Understanding the mechanisms that underlie the acceleration of electrons in the lower solar corona and their subsequent access to interplanetary space is essential to answering unsolved questions in heliophysics that form part of the main science questions of the Solar Orbiter and Parker Solar Probe missions. Plasma ejections are continuously observed in different forms, such as coronal mass ejections (CMEs), filament eruptions, or plasma jets. Among all these phenomena, plasma jets, defined as collimated plasma beams \citep[for a review, see][]{2016SSRv..201....1R}, are of a particular interest since they are ubiquitous on the Sun. Indeed, they can be observed in active regions \citep[ARs; e.g.,][]{2004ApJ...604..442L,2009A&A...508.1443B,2011ApJ...742...82K,Glesener_2012,2022A&A...665A..29O}, coronal holes \citep[CHs; e.g.,][]{2007PASJ...59S.771S,2010A&A...516A..50S,2013ApJ...775...22S} or even in the quiet Sun \citep[e.g., ][]{Hou_2021}. Because of the nature of the jets, they are thought to be an important contributor in continuously supplying mass and energy to the upper solar atmosphere and being at the origin of solar winds \citep[e.g.,][]{2011Sci...331...55D}.

Coronal jets, which are plasma jets occurring in the corona, can be associated with solar flares and they can be observed in extreme ultraviolet (EUV) as well as in X-rays \citep[e.g.,][]{2011ApJ...742...82K,Glesener_2012,2020ApJ...889..183M}. Despite coronal jets being widely observed, their formation mechanism is still under debate. Indeed, in  recent decades, different triggering processes have been proposed. Initially, the interchange reconnection model based on emerging flux came into the limelight \citep{1977ApJ...216..123H,1989ApJ...345..584S,1992PASJ...44L.173S,1996PASJ...48..353Y}, in which open magnetic field lines reconnect with closed, emerging magnetic field lines. Herein, the hot jet, which has a temperature of the order of several MK, results from a fast shock produced near the reconnection site. 
As well as a hot jet, chromospheric ejections can be observed as cool jets or surges, often observed in H$\alpha$ \citep{1996ApJ...464.1016C}.
More recently, in contrast to the emerging flux model, minifilament eruption events have been proposed as drivers of the jet-producing reconnection \citep[e.g., ][]{2010ApJ...720..757M,2015Natur.523..437S}. Minifilament eruptions are the small scale versions of filament eruptions that initiate CMEs. 

Hard X-ray (HXR) observations are of particular interest for investigating coronal jets associated with solar flares since they give a direct diagnosis of the acceleration of high-energy electrons via bremsstrahlung emissions. Magnetic reconnection, which releases free magnetic energy into
various other forms of energy, is also responsible for the acceleration of high-energy electrons. Some accelerated electrons heat the ambient plasma to temperatures on the order of tens of MK and other electrons can escape along open magnetic field lines \citep[for a review, see][]{2011SSRv..159...19F,2017LRSP...14....2B}. 
Another interesting aspect of HXR observations is that it is possible to diagnose the efficiency of the mechanism for accelerating particles in terms of the energy transferred to nonthermal electrons. 
While microflares typically show steep HXR spectra, indicating that they are less efficient in accelerating high-energy electrons than larger flares \cite[e.g.,][]{Stoiser2007,2008ApJ...677..704H,2014ApJ...789..116I,2016A&A...588A.116W},
there are also microflare observations showing significantly prominent nonthermal emission \citep[e.g., ][]{2008A&A...481L..45H,2013ApJ...765..143I}. This suggests that (other than the dependency on flare size) there may be other factors that significantly influence the acceleration efficiency. Over the past two decades, several studies reported on the occurrence of flares exhibiting exceptionally prominent nonthermal emissions \citep{1997A&A...320..620F,2007ApJ...670..862S,Lysenko_2018}, referred to as "early impulsive" flares. However, detailed studies on the morphologies of these events investigating the mechanism at the origin of the phenomenon are still lacking.

We report on the observations of the SOL2021-02-18T18:05 microflare of GOES A8 class above the pre-flare background that is associated with a coronal jet and prominent HXR nonthermal emission. In the following (Sect.~\ref{sec:obs-analysis}), we describe the data analysis and the multi-instrument observations. The discussions of the observations in the framework of the interchange reconnection scenario are presented in Sect.~\ref{sec:discussions}. We give our summary and conclusions in Sect.~\ref{sec:conclusions}


\section{Observations and data analysis \label{sec:obs-analysis}}

The Spectrometer/Telescope for Imaging X-rays \citep[STIX, ][]{2020A&A...642A..15K} 
aboard the Solar Orbiter spacecraft \citep{2020A&A...642A...1M} 
is designed to observe a wide range of solar flares in the energy range from 4 to 150 keV. While the diagnostic capabilities of STIX resemble those of its predecessor, namely, Reuven Ramaty High-Energy Solar Spectroscopic Imager \citep[RHESSI,][]{2002SoPh..210....3L}, one of the advantages of STIX is the constant non-solar background during flaring time-scales. This implies a more simplified detection and analysis of small events \citep{2021A&A...656A...4B,Saqri2022}, such as the microflare studied in the present work.
Solar Orbiter's deep-space trajectory results in STIX having a different view point of the Sun with respect to Earth, most of the time. During the observation of the SOL2021-02-18T18:05 microflare of GOES A8 class, Solar Orbiter was at a distance of 0.51 AU from the Sun with a separation angle to the Sun-Earth line of about 149$^\circ$ East (see Fig.~\ref{fig:diff-look-directions}). The different distance to the Sun of Solar Orbiter relative to Earth implies a different photon arrival time. Consequently, for all figures shown in this paper, the STIX times have been corrected by +239.9 s to take into account the shorter light travel time from the flare site to Solar Orbiter compared to Earth. As seen from Earth, the microflare was located close to the eastern limb, whereas from the Solar Orbiter vantage point, it was seen close to the western limb (bottom row of Fig.~\ref{fig:diff-look-directions}).

In addition to STIX observations, we also use data from the Extreme Ultraviolet Imager \citep[EUI;][]{2020A&A...642A...8R} on board Solar Orbiter, the Solar X-ray Monitor \citep[XSM;][]{2020CSci..118...45S} on board the Chandrayaan-2 \citep{2020LPI....51.1994V} satellite, the Atmospheric Imaging Assembly \citep[AIA;][]{2012SoPh..275...17L} on board the Solar Dynamics Observatory \citep[SDO;][]{2012SoPh..275....3P}, the Extreme UltraViolet Imager \citep[EUVI;][]{2008SSRv..136...67H} on board the Solar Terrestrial Relations Observatory Ahead \citep[STEREO-A;][]{2008SSRv..136....5K}, the X-ray Sensor \citep[XRS;][]{1996SPIE.2812..344H} of the Geostationary Operational Environmental Satellite (GOES), the Greenland station of the Compound Astronomical Low frequency Low cost Instrument for Spectroscopy and Transportable Observatory \citep[CALLISTO;][]{2009EM&P..104..277B}, and the telescope in Cerro Tololo, Chile, of the Global Oscillation Network Group \citep[GONG;][]{1996Sci...272.1284H}. 
In order to obtain the XSM time profiles at the nominal 1 s cadence, we used the XSM Data Analysis Software \citep[XSMDAS;][]{2021A&C....3400449M} and we subsequently integrated them to reduce noise.
We used L2 data products of the EUI/FSI (Full Sun Imager) 174 \AA{} data, which were de-rotated by the roll angle in order to obtain the final image with solar north up. 
The XRS data are from GOES-17 and have been obtained through the GOES workbench within SSWIDL. 
The EUVI and AIA data have been calibrated using the standard software in SSWIDL \textit{secchi\_prep.pro} and \textit{aia\_prep.pro}, respectively. 
The AIA time profiles shown in Sect.~\ref{sec:time-profiles} have been obtained by spatially integrating around the flare region only, excluding emission due to the propagating jet. The STIX light curves were instead derived from full-disk measurements.
The GONG and CALLISTO data are available as standard data products, therefore, they have been applied as such.
    
    \subsection{Timing analysis \label{sec:time-profiles}}

\begin{figure}
    \centering
    \includegraphics[width=0.44\textwidth]{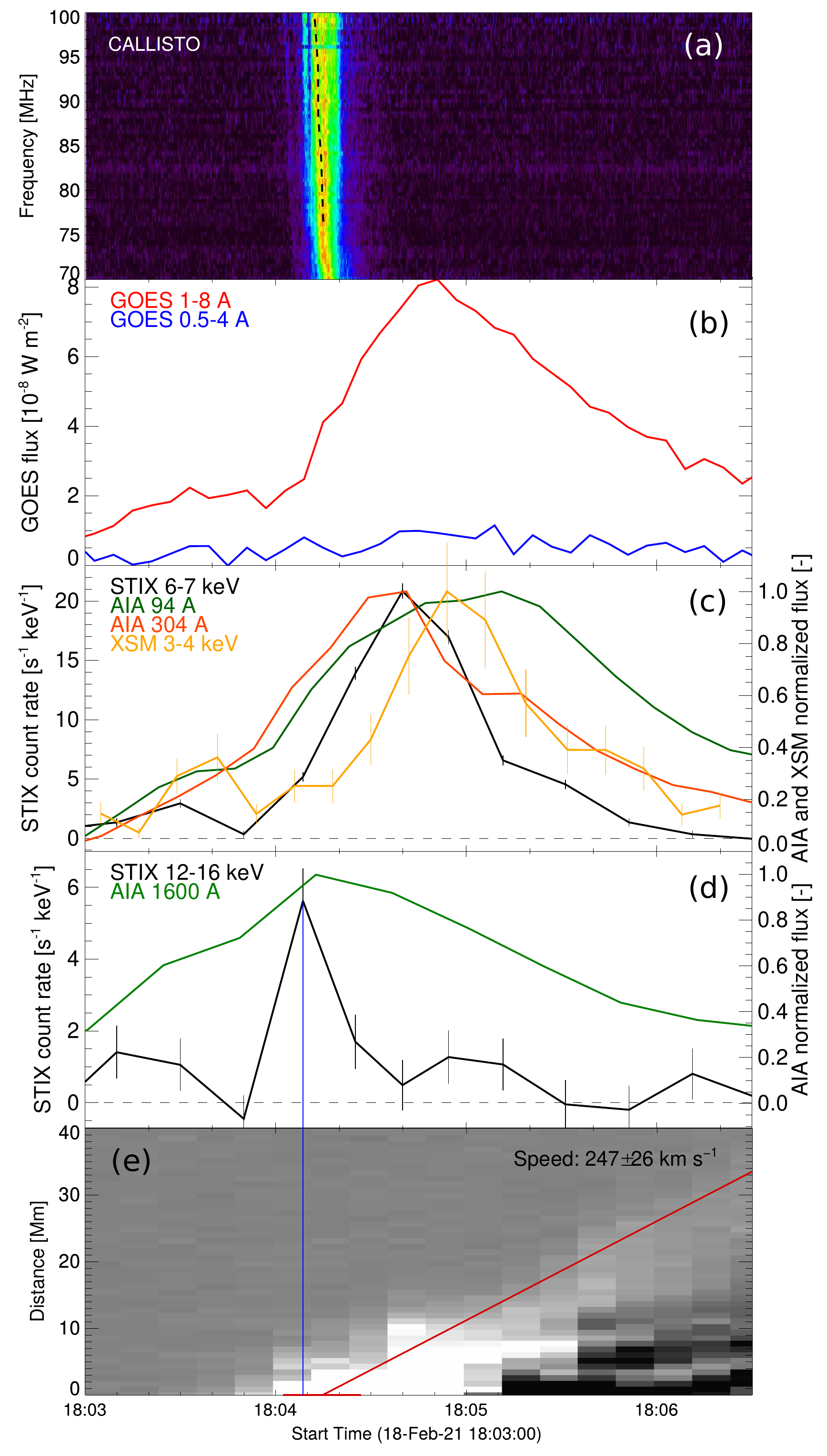}
    \caption{
    Time profiles of the event emissions at different wavelengths. Type III radio bursts observations of the CALLISTO station in Kangerlussuaq, Greenland, are given in panel (\emph{a}), where the black dashed line represents the profile used to derive the drift rate. GOES/XRS temporal profiles are represented in panel (\emph{b}). Panel (\emph{c}) includes the lightcurves of the thermal emission observed by STIX (6-7 keV, black) and XSM (3-4 keV, orange), whereas in green and red the lightcurves of the AIA 94 \AA{} and 304 \AA{}, respectively. Panel (\emph{d}) shows the nonthermal emission detected by STIX (12-16 keV, black) and the emission mostly coming from the chromosphere observable in the AIA 1600 \AA{} band (light green). Finally, panel (\emph{e}) represents the distance-time plot of the initial phase of the jet obtained from the running difference of the AIA 304 \AA{} maps. The red line indicates the peak emission
    of the outward moving jet, corresponding to a speed in the plane-of-sky of $247 \pm 26 ~\textrm{km}~\textrm{s}^{-1}$.
    The vertical blue line, which indicates the peak time of the STIX nonthermal emission, nicely correlates with the origin of the jet that is indicated by the horizontal red segment at $y=0$. }
    \label{fig:time-profiles}
\end{figure}

In order to understand the overall evolution of the microflare and the related coronal jet,  the time profiles at different wavelengths are presented in Fig.~\ref{fig:time-profiles}.
The dynamic radio spectrum in panel (\emph{a}) shows type III radio bursts, which are a signature of propagating nonthermal electron beams that generate Langmuir waves at the local plasma frequency \citep[for a review, see][]{2014RAA....14..773R}. The timing of the type III radio bursts correlates with the peak time that is observed in the STIX time profile of the 12-16 keV energy band in panel (\emph{d}). This motivates the interpretation of the presence of a nonthermal component in the higher energies of STIX.
The black dashed line indicates the profile used to deduce the drift rate, which is on the order of $\sim 20~\mathrm{MHz}~\mathrm{s}^{-1}$. Assuming the Newkirk density model profile \citep{1961ApJ...133..983N}, 
where we accounted for the enhanced densities within ARs with respect to the quiet Sun,
we derived the speed of the radio burst source. For a burst observed at fundamental plasma emission, we estimated the upwardly directed speed to be of the order of $\sim 70'000 - 90'000~\mathrm{km}~\mathrm{s}^{-1}$ (or $\sim 0.23 - 0.30~\mathrm{c}$), 
which corresponds respectively to about 14 and 24 keV in electron energies, and the emission height to range from $\sim 0.3$ (at $100~\mathrm{MHz}$) to $\sim 0.8$ (at $70~\mathrm{MHz}$) \(\textup{R}_\odot\) above the Sun surface. This confirms the type III nature of the radio burst.
It is worth mentioning that the type III burst is seen to continue into interplanetary space, up to about $0.1\,\mathrm{MHz}$, as observed by the Radio and Plasma Wave Investigation \citep[WAVES;][]{1995SSRv...71..231B} experiment aboard the WIND spacecraft. This confirms that the accelerated electrons escape into interplanetary space.
No CME was reported by the LASCO CME catalog.

Another interesting aspect of the nonthermal observation resides in the similar peak time of the AIA 1600 \AA{} emission with the STIX 12-16 keV time profiles. In the standard flare scenario, this is consistent with the accelerated electrons colliding with the much denser chromosphere. The interpretation of the slower decay of the AIA 1600 \AA{} light curve with respect to STIX hard X-rays is that the decay of the AIA 1600 \AA{} emission is dominated by the cooling of the plasma, whereas for STIX, as soon as the electron acceleration process and the almost instantaneous interaction with the ambient protons (producing nonthermal bremsstrahlung) ceases, the 12-16 keV light curve drops.

The time evolution of the microflare associated with the jet and the type III radio burst is shown in panel (\emph{b}) with the GOES/XRS lightcurves and in panel (\emph{c}) with STIX 6-7 keV, XSM and SDO/AIA. The lack of signal in the GOES 0.5-4 \AA{} band may be due to the relatively high background during the flare interval. The difference in the peak time between the STIX low energy band, GOES and AIA is due to the sensitivity of the instruments to different plasma temperatures \citep{2021A&A...656A...4B}.
Interestingly, all time profiles show pre-flare plasma heating (at about 18:03:30 UT), which indicates that the energy dissipation already started before the main nonthermal peak observed by STIX in the 12-16 keV emission profile.

In order to obtain the initial jet speed (overview of the jet in Fig.\,\ref{fig:diff-look-directions}) in the plane of the sky and estimate its onset time, we constructed a distance-time plot. To do so, we extracted the intensity along a straight line on the running difference maps (the location of the straight line is represented by the dashed segment 
in Fig.~\ref{fig:flare_20210218}) and stacked together the intensities at different times. The distance-time plot shown in panel (\emph{e}) has been obtained with AIA 304 \AA{} running difference maps. 
The red line shows the linear fit to the peak emission of the outward moving jet and its slope corresponds to a linear speed of the jet in the SDO plane-of-sky of $247 \pm 26 ~\textrm{km}~\textrm{s}^{-1}$.
The same analysis has been done with the AIA 171 and 193 \AA{} maps, where the signal of the propagating jet is clearly visible. The obtained results are similar to the value reported in panel (\emph{e}) and the average speed among all these wavelengths is $251 \pm 59 ~\textrm{km}~\textrm{s}^{-1}$. The confidence interval of the onset time, which is calculated considering $1-\sigma$ uncertainty in the slope (i.e., the uncertainty of the speed), is represented with the red horizontal line at $y=0$. We can observe that the confidence interval is consistent with the peak time in the STIX nonthermal profile. This suggests that particle acceleration and the creation of the jet are closely linked.


    \subsection{UV/EUV imaging analysis \label{subsec:EUVandUVimaging}}
    
\begin{figure*} 
    \centering
    \includegraphics[width=0.85\textwidth]{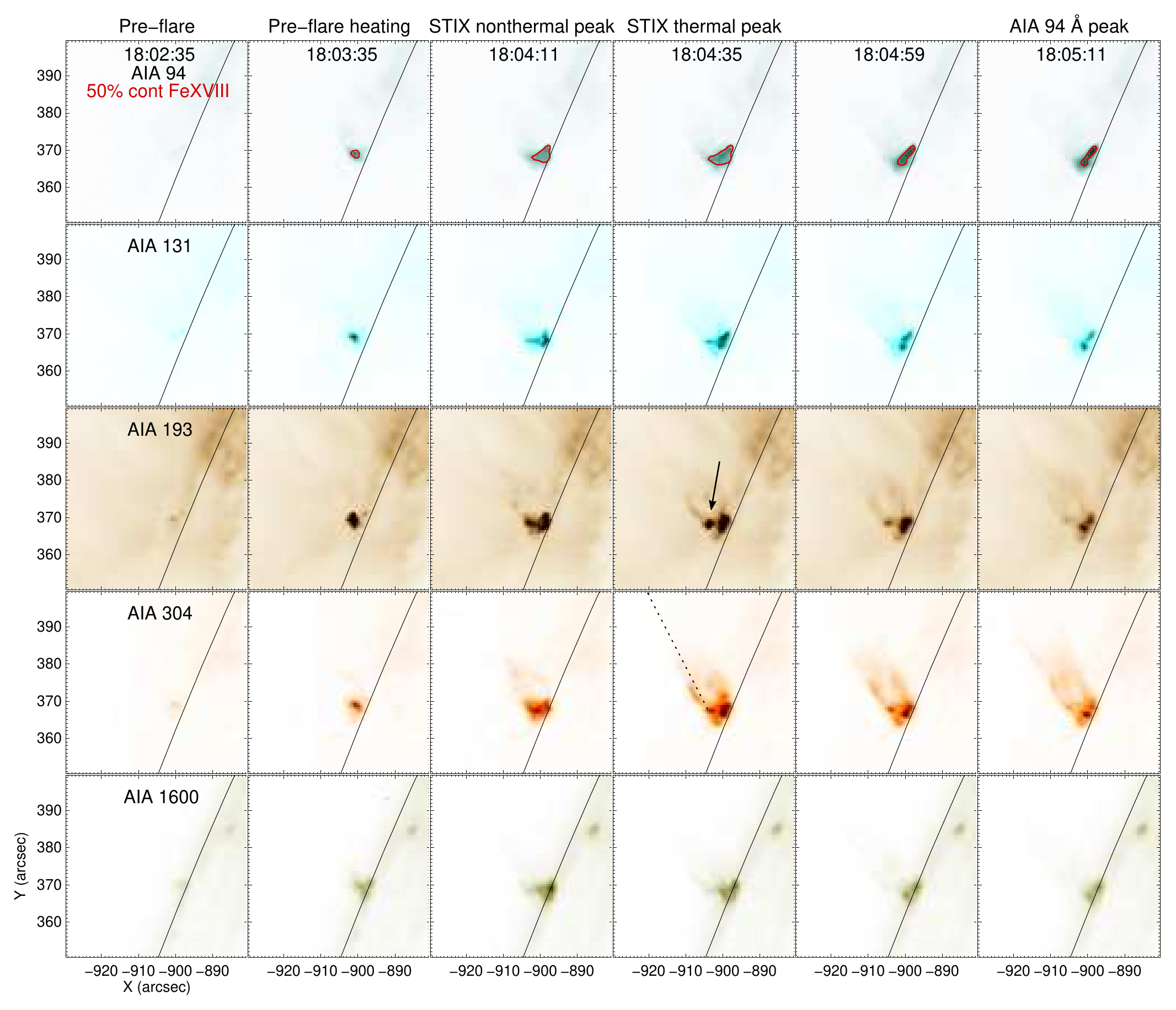}
    \caption{SDO/AIA image sequences at different wavelengths for the SOL2021-02-18T18:05 event. From top to bottom: AIA 94 \AA{} maps with in red the 50\% contour of the \ion{Fe}{xviii} line complex maps \citep{2013A&A...558A..73D}, AIA 131 \AA{}, AIA 193 \AA{}, AIA 304 \AA{} and AIA 1600 \AA{}. The black dashed segment in the AIA 304 \AA{} map represents the line that has been used for constructing the distance-time plot in Fig.~\ref{fig:time-profiles}. The black arrow indicates the above-the-loop location coinciding with the base of the jet and the electron acceleration site.
    }
    \label{fig:flare_20210218}
\end{figure*} 

The overall evolution of the EUV and UV images obtained with the SDO/AIA telescope is depicted in Fig.~\ref{fig:flare_20210218}, where (as seen from Earth) the event occurs near the eastern limb. From left to right, we show the time evolution of the event, whereas from top to bottom we show the different selected wavelengths.
Initially, from the first column, which represents a time frame approximately two minutes before the STIX nonthermal peak, no clear flare-related signal is observable in any of the maps. Afterwards, about 45 seconds before the STIX nonthermal peak, pre-flare heating occurs. This is consistent with the time profiles in Fig.~\ref{fig:time-profiles}, where signal is detected in the XSM, AIA, GOES and both STIX lightcurves before the main nonthermal peak of the STIX 12-16 keV.

At the time of the STIX nonthermal peak, the AIA 1600 \AA{} map shows the peak intensity on-disk, most likely coinciding with chromospheric heights. This is in agreement with the accelerated electrons depositing energy in the lower part of the atmosphere. However, since the event is observed near the limb, some of the emission could be blocked by absorbing foreground plasma. All other wavelengths reveal emission from higher altitudes indicating that the chromospheric heated plasma has expanded along the magnetic field lines. Since the EUV limb extends higher up with respect to the UV limb, it is expected that most of the chromospheric EUV emission is occulted. 
In addition, we note that at this time there is a secondary region peaking in the AIA 1600 \AA{} map, at $(x,y) \sim (-885'',385'')$. A possible interpretation may be that some upwardly accelerated electrons gained access to open field lines and escaped
towards interplanetary space (see Sect. \ref{sec:time-profiles}), whereas some upwardly propagating electrons turned back toward the solar surface due to the propagation along closed field lines. The latter would explain the appearance of this secondary source in the AIA 1600 \AA{} maps. The  magenta dot pointed by the arrow in Fig.\,\ref{fig:diff-look-directions} indicate the location of this remote enhancement observed with the 171 \AA{} passband of SDO/AIA.

After the STIX thermal peak occurring at 18:04:40 UT, the top of the flare loop can be observed in the AIA 94 \AA{} map. Again, here the occultation is likely to play a role, blocking the emission coming from the lower part of the loop. Since the AIA 94 \AA{} filter has two temperature peaks, at $\sim 1\,\mathrm{MK}$ and at $\sim 7\,\mathrm{MK}$, we extracted the emission from the hotter peak by approximating the \ion{Fe}{xviii} emission, as described in \citet{2013A&A...558A..73D}.
The fact that the 50\% contours of the \ion{Fe}{xviii} line complex coincide with the 94 \AA{} emission indicates that the temperature of the plasma mostly originates from the hotter peak in the response of the 94 \AA{} band. 
Similarly, the loop-top can be observed in the 131 \AA{}, 193 \AA{} and 304 \AA{} maps.

\begin{figure*}
    \centering
    \includegraphics[width=0.9\textwidth]{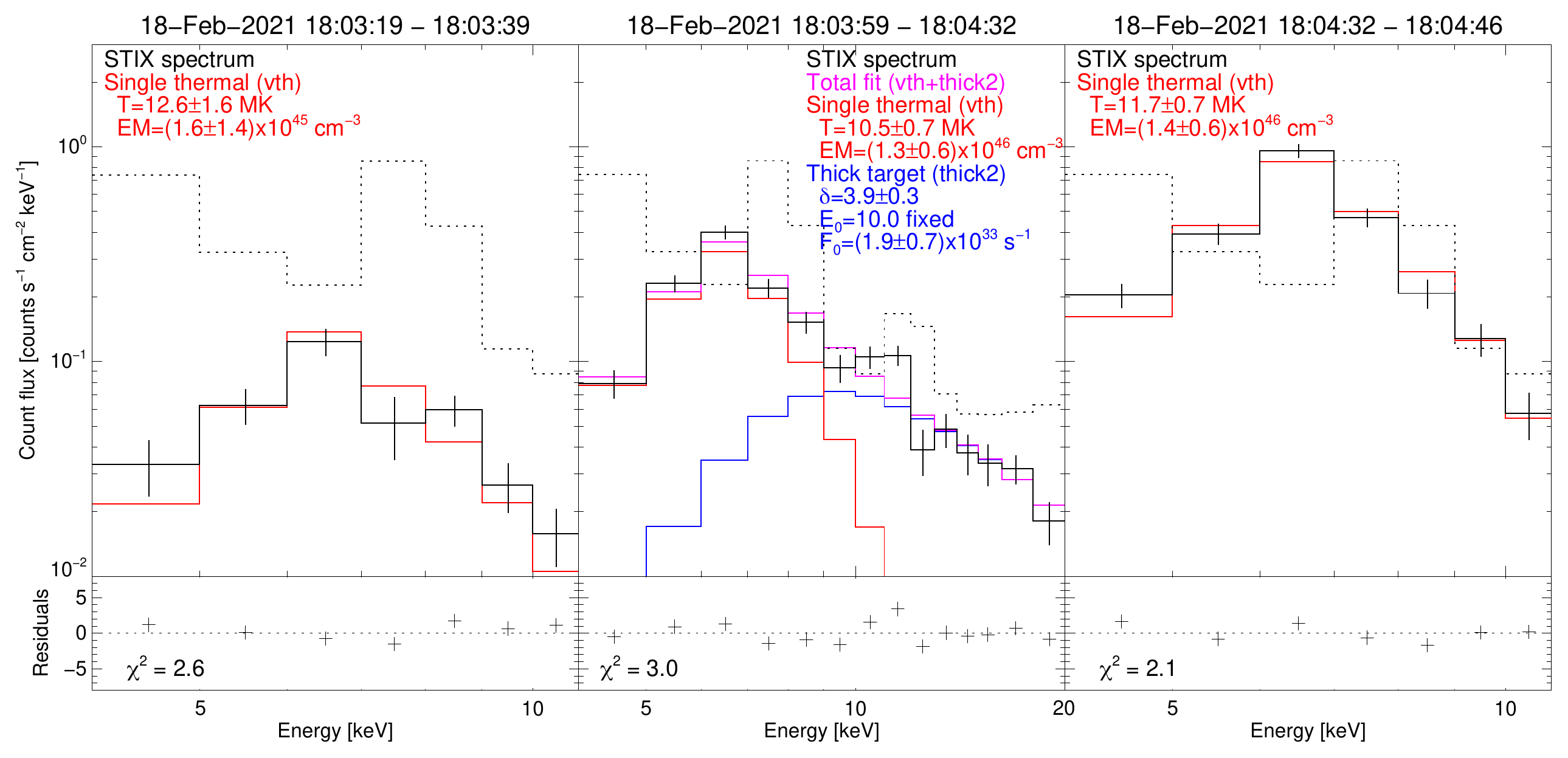}
    \caption{Solar Orbiter/STIX background-subtracted count spectra (solid black) at different times during the flare. 
    The STIX spectrum during the pre-flare phase (\emph{left})
    can be fitted by assuming an isothermal model (red).
    Around the STIX nonthermal peak (\emph{center}), the spectrum can be fitted with an isothermal and a thick target model (blue), whereas around the thermal peak (\emph{right}), we assumed an isothermal model only. The dashed black curves in each plot represent the STIX background spectra taken during non-flaring times close to the event. Below each plot  we report the residuals, observations minus total fit, in units of the standard deviation calculated from counting statistics. The resulting fit parameters are shown in the legend of each plot.}
    \label{fig:spectra}
\end{figure*}

The most interesting (and never before reported) feature in the EUV observation is indicated by the black arrow. First of all, this corresponds to a location that lies above the top of the flare loop. 
Secondly, the coronal jet originates from this location and, indeed, it has been chosen to be the origin of the distance-time plot shown in Fig.~\ref{fig:time-profiles}. Moreover, this source is already present in the pre-flare heating phase and  moves away from the solar surface in the course of time, with an average projected bulk speed of about $30\,\textrm{km}\,\textrm{s}^{-1}$. In the magnetic reconnection scenario, this is consistent with emerging closed field lines reconnecting with open field lines or large-scale loops (see Sect.\,\ref{sec:discussions}), at higher altitudes in the course of the time. 
Therefore, the source most likely outlines the location of the magnetic reconnection region. 

A word of caution is in order here. We see no hints in the observations of the presence of a small-scale erupting filament as the trigger of the magnetic reconnection, as described by the mini-CME scenario \citep[e.g.,][]{2010ApJ...720..757M,2015Natur.523..437S}. This may be due to unresolved structures.


   \subsection{Spectroscopic analysis \label{subsec:spectroscopy}}
    
\begin{figure*}[h]
    \centering
    \includegraphics[width=0.94\textwidth]{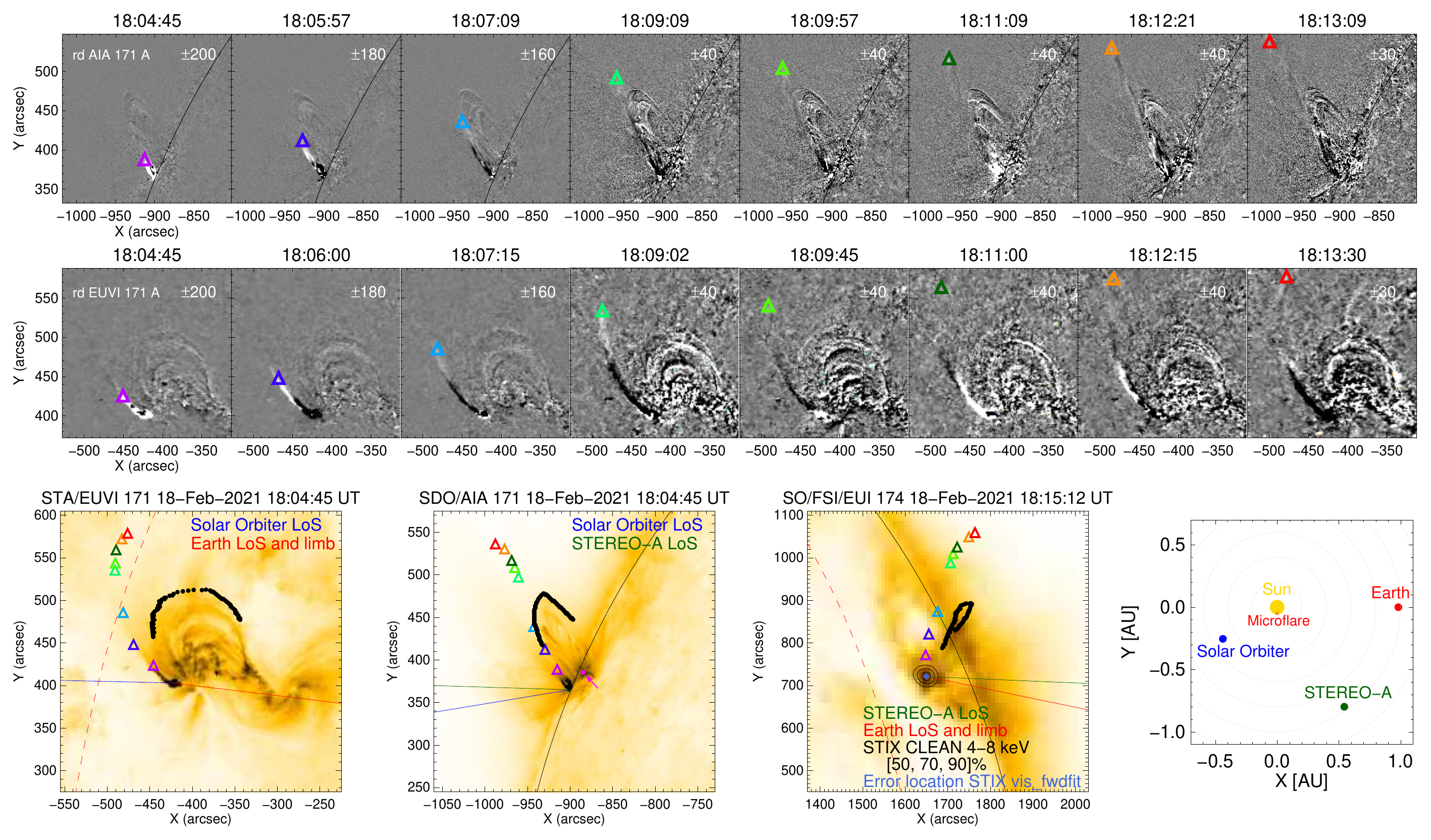}
    \caption{3D reconstruction of the propagating coronal jet. The top and middle row show the running difference maps of the SDO/AIA 171 \AA{} and the STEREO-A/EUVI 171 \AA{} maps, respectively. The colored triangles indicate the same features seen from the two look directions for different times. The bottom row includes three EUV maps and one insert: from left to right, respectively, the maps of STEREO-A/EUVI 171 \AA{}, SDO/AIA 171 \AA{} and Solar Orbiter/EUI 174 \AA{}, and a scheme depicting the locations of the three spacecrafts. On top of the EUV images, the same colored triangles indicate the trajectory of the coronal jet and the black dots represent the reconstruction of the top part of the overarching loop. The solid lines represent the different line of sights (LoS) as seen from each individual vantage point.  The black contours on top of the EUI map show the STIX thermal image. The magenta dot pointed by the arrow on top of the SDO/AIA map indicates the location of the remote enhancement.
    }
    \label{fig:diff-look-directions}
\end{figure*}

The spectroscopic analysis for this small STIX flare is challenging. The detected flare signal is fainter than the instrumental background and contains only slightly over a thousand counts in total. The STIX science data used for the spectral fitting are the compressed pixel data \citep{2020A&A...642A..15K} that allows for detailed corrections to be applied to each pixel and detector separately. The software to convert the STIX data to a format readable by OSPEX is still under development and the version used here is from March 2022.  In order to account for currently unknown systematic effects in the calibration, we assumed the existence of an additional source of error and added a  5\% systematic uncertainty in quadrature to the errors from photon counting statistics. 

In Fig.~\ref{fig:spectra}, we present a spectral fitting during the three phases, the pre-flare phase, the impulsive phase, and the time of the peak of the thermal emission.

During the pre-flare phase, counting statistics are extremely low, and the error bars are very large, accordingly. Nevertheless, a thermal fit gives a temperature of $12.6 \pm 1.6~\mathrm{MK}$ and an EM of $(1.6 \pm 1.4) \times 10^{45}~\mathrm{cm}^{-3}$. The derived temperature is consistent with the detection of the pre-flare source in the AIA 131 \AA{} image.

Around the nonthermal peak, different fit models have been checked. A purely isothermal fit does not represent the data reasonably well ($\chi^2 = 14.7$) and the resulting temperature is very hot ($\sim 36 \, \mathrm{MK}$). Using a double thermal fit, instead, results in a much better fit ($\chi^2 = 2.8$), but the temperature of the hotter component is extremely high and not plausible for an event of this size ($\sim 79\, \mathrm{MK}$, i.e., an average electron energy of $\sim 7$ keV). An isothermal and standard thick-target fit model represents the data equally well ($\chi^2 = 3.0$) and gives more reasonable parameters, as shown in the following. Moreover, this model is consistent with the time correlation of the STIX higher energies with type III radio bursts, which are signatures of propagating electron beams (see Sect.~\ref{sec:time-profiles}). This motivates the interpretation of the presence of a nonthermal component in the higher energies of STIX. Such a model results in a temperature of $10.5 \pm 0.7~\mathrm{MK}$ and an EM of $(1.3 \pm 0.6) \times 10^{46}~\mathrm{cm}^{-3}$. We find that counts above 10 keV are mainly nonthermal and that the electron spectral index corresponds to $\delta = 3.9 \pm  0.3$, which in the photon space corresponds to $\gamma = 2.9 \pm  0.3$. Compared to the statistical microflare study by \citet{2008ApJ...677..704H}, which shows a median photon index of 6.9, the presented slope is extremely flat, but not outside the previously published range. This microflare compares to the hard microflares reported by \cite{2008A&A...481L..45H} and \cite{Ishikawa_2013}, indicating that the process of accelerating particles in this event was particularly efficient. The low-energy cutoff in the thick target fit is only very poorly constrained, and the nonthermal energy content was thus calculated for a cutoff fixed at 10 keV. Consequently, the resulting electron flux of $F_0 = (1.9 \pm 0.7) \times 10^{33}~\mathrm{s}^{-1}$ has to be interpreted as a lower limit. The resulting lower limit of the nonthermal energy, obtained by integrating the nonthermal power over 33 s, namely, the duration of the nonthermal spectral interval, corresponds to $(1.6 \pm 0.5)\times 10^{27}$ erg. For the sake of completeness, we also tried to fit the spectrum with the warm thick-target fit and a similar power law and $\chi^2$ are obtained. However, the errors on the fitted warm target parameters are huge, that is, larger than the values themselves, since we are trying to constrain six parameters with 14 energy bins and with limited statistics.
    
Around the STIX thermal peak, an isothermal model with a temperature of $11.7 \pm 0.7~\mathrm{MK}$ and an EM of $(1.4 \pm 0.6) \times 10^{46}~\mathrm{cm}^{-3}$ fits the data satisfactory, without the need for a nonthermal component. 

As the errors on the fit thermal parameters are rather large, it is difficult to draw any meaningful conclusions on the temperature evolution in time. However, it is clear that pre-flare heating is significant since it heats plasma to roughly the same temperature as during the flare, but at a lower EM.


    \subsection{3D properties of the propagating jet}

By taking advantage of the different viewpoint of STEREO-A with respect to SDO, we combined the multi-vantage point observations to derive the 3D reconstructions of the propagating jet path and heights as well as its velocity using the tie-pointing and triangulation techniques \citep{THOMPSON2009351, Inhester2006}. We identified identical features in SDO and STEREO-A images and used an algorithm of 3D reconstructions based on epipolar geometry, as described in detail in \citet{Podladchikova2019}.

Figure~\ref{fig:diff-look-directions} shows a sequence of AIA 171 \AA{} (\emph{top}) and STEREO-A/EUVI 171 \AA{} (\emph{middle}) running difference maps, where the colored triangles highlight the propagation of the coronal jet based on matching the same features on both AIA and EUVI images. The same triangles are plotted on top of three EUV maps (\emph{bottom}), from left to right: STEREO-A/EUVI 171 \AA{}, SDO/AIA 171 \AA,{} and Solar Orbiter/EUI 174 \AA{}, respectively. The bottom-right panel depicts the location of the three spacecraft at the time of the flare. The STIX reconstructed image has been over-plotted on the EUI map. For more details, we refer to Sect.\,\ref{sec:STIX-imaging}.
    
\begin{figure}
    \centering
    \includegraphics[width=0.42\textwidth]{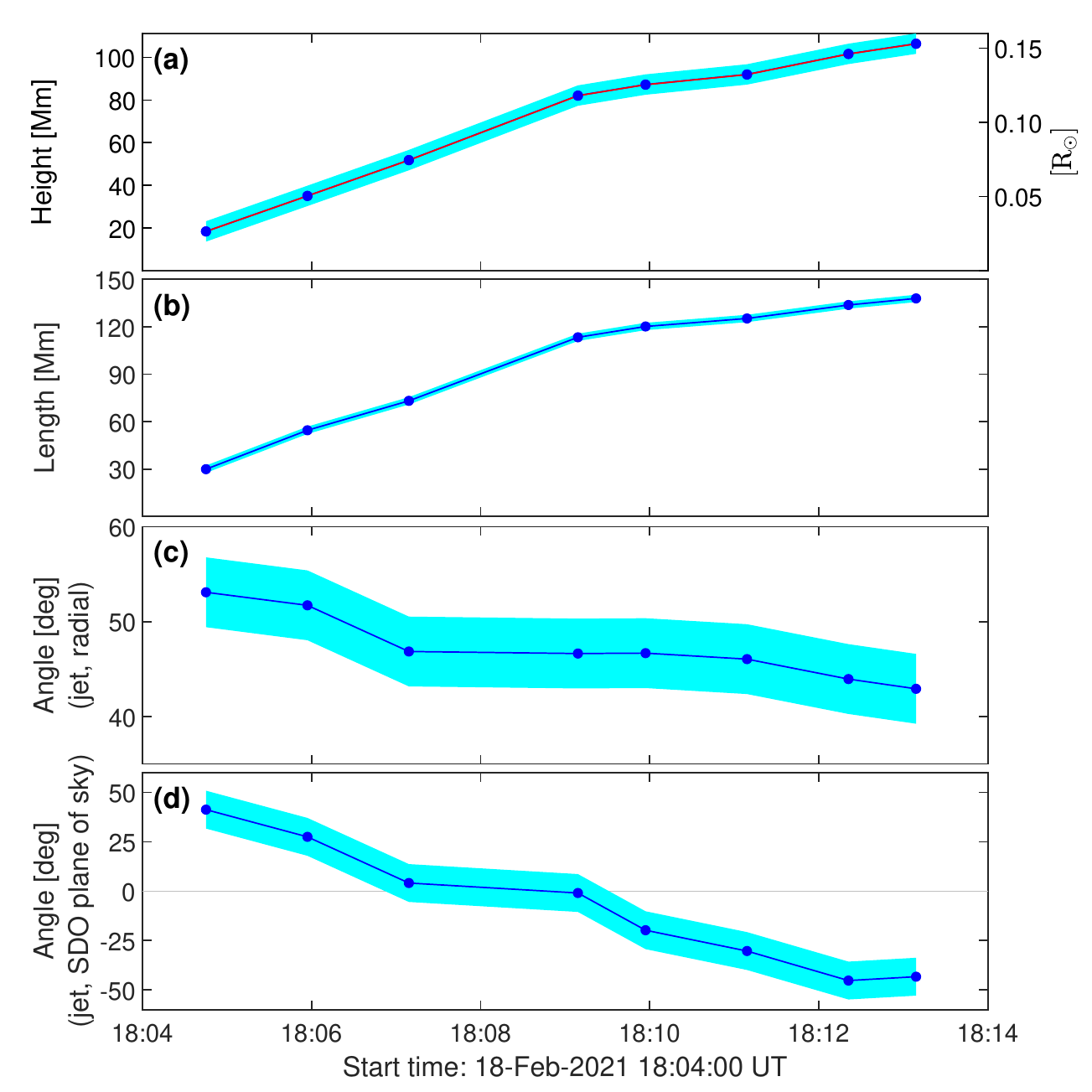}
    \caption{Jet characteristics identified from the 3D reconstructions: (\emph{a}) Evolution of the jet height above the solar surface. (\emph{b})  Jet length along its trajectory (distance traveled). (\emph{c})  Angle between the jet front and the radial direction. (\emph{d}) Angle between the vector formed from the jet front at time, $t_i$, to the front at the following time step, $t_{i+1}$, and the SDO plane-of-sky, showing the change of direction of the jet trajectory with respect to the SDO viewpoint. Shaded areas show $3\sigma$ errors by shifting the selected points in SDO images by 1 pixel.}
    \label{fig:3D-height-angle}
\end{figure}

What is striking about the visual representation of the jet trajectory is how different it appears from the different viewpoints. As seen from SDO, the collimated plasma beam appears to occur behind the loop and roughly along a straight line, whereas when it is observed by STEREO-A and Solar Orbiter, its trajectory is clearly curved. This highlights the significant influence of a certain viewpoint when analyzing 3D structures with 2D images. 
To gain a better perception of the trajectory of the propagating jet, Fig.~\ref{fig:3D-height-angle} reports the jet characteristics identified from the 3D reconstructions. Panel~(\emph{a}) shows the evolution of the jet height above the solar surface, which sharply rises from around 18 to 82 Mm from 18:04:45 to 18:09:09~UT (264 seconds). Data taken at later times do not allow us to trace the leading edge of the jet reliably; thus, the corresponding estimations at times from 18:09:57~UT to 18:13:09 reflect the increase of inner parts of the jet from 87 to 106 Mm.

Panel~(\emph{b}) in Fig.~\ref{fig:3D-height-angle} shows the estimated length of the jet determined from the base to the topmost part, along the jet trajectory. From 18:04:45 to 18:09:09~UT the jet length rises from around 30 to 113~Mm with a further gradual increase to 138~Mm reached at 18:13~UT. Starting around 18:09~UT the intensity and signal-to-noise ratio of the jet front becomes weaker. As a consequence, at later stages, we might not identify the outermost part of the jet front, but parts that are located further inner of the jet instead. Therefore, at these later times the derived heights, lengths, and velocities may only be lower estimates of the jet front kinematics. For this reason, we derived the mean velocity of the jet only over the time range 18:04:45 to 18:09:09~UT, where the jet front is identified clearly in the different instruments and the measurement values are thus more certain. We obtained the 3D velocity of the jet as $v \approx 312 \pm 40\, \textrm{km}\, \textrm{s}^{-1}$, which is faster than the speed in the plane of sky deduced in Fig.\ref{fig:time-profiles}. This is consistent with the fact that the projected 2D speed is only a lower limit for the true 3D velocity \citep{Podladchikova2019}.

Panel~(\emph{c}) in Fig.~\ref{fig:3D-height-angle} shows the angle between the jet front and the radial direction determined at the base of the jet. As can be seen, during the jet evolution, the angle between the jet trajectory and the radial direction is continuously changing, indicating that the jet is not moving along a straight line.
In panel~(\emph{d}), we visualize the change of direction of the jet trajectory with respect to the SDO viewpoint 
by calculating at each time step the angle between the vector formed from the jet front at time $t_i$ to the front at the following time step $t_{i+1}$ 
and the SDO plane of sky.
It is seen that the angle between the local vector of the jet trajectory and SDO's plane of sky changes steadily from $+41^{\circ}$ at 18:04~UT to $-43^{\circ}$ at 18:13 UT.
This demonstrates that the jet is first moving outward across SDO's plane-of-sky and then changing to moving inward. These findings suggest that the jet moves along a curved trajectory across SDO's plane of sky. This behavior is not evident from the SDO observations in Fig. 4, as the curvature of the jet trajectory (which is clearly seen from the STEREO and Solar Orbiter view) is mostly perpendicular to the SDO plane of sky. However, as the highest measured point of the trajectory is still increasing in altitude, and factoring in that type III radio bursts indicate that some energetic electrons have escaped the Sun, it is most likely that  the jet also escapes. 

\begin{figure*}[h]
    \centering
    \includegraphics[width=\textwidth]{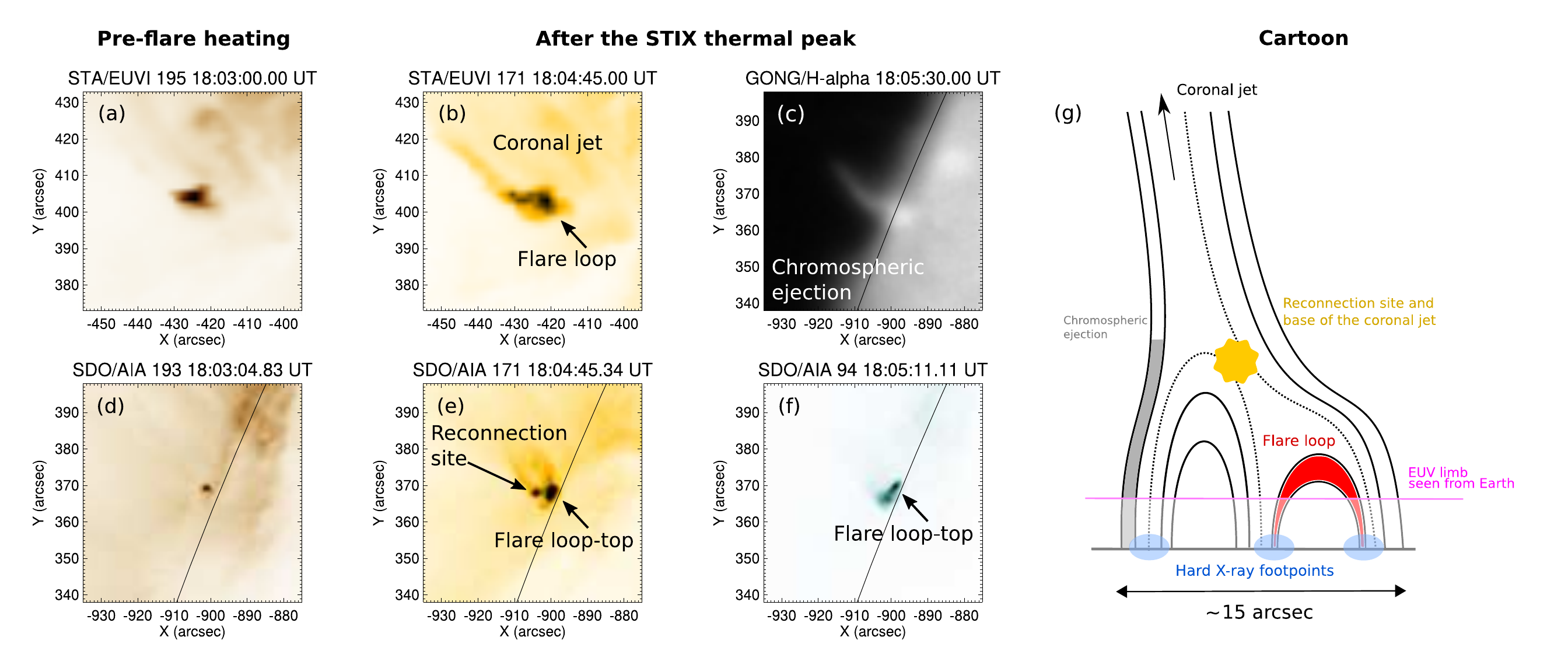}
    \caption{Summary plot with observations and cartoon of the SOL2021-02-18T18:05 event. The leftmost column (panels (\emph{a}) and (\emph{d})) shows the SDO/AIA 193 \AA{} and STEREO-A/EUVI 195 \AA{} maps during the pre-heating phase. The second and third columns (panels (\emph{b}), (\emph{c}), (\emph{e}), and (\emph{f})) show, respectively, STEREO-A/EUVI 171 \AA{}, GONG/H$\alpha$, SDO/AIA 171 \AA,{} and 94 \AA{} maps after the STIX thermal peak. The drawing in the rightmost panel
    depicts the cartoon of the interchange reconnection based on the emerging flux model in the case of the observed event.}
    \label{fig:cartoon}
\end{figure*}


    \subsection{STIX HXR imaging \label{sec:STIX-imaging}}

In the following, we introduce the STIX HXR imaging reconstruction of the microflare under investigation. For this event, we present an image in the thermal energy range only, since a robust nonthermal image reconstruction is not possible due to low counting statistics at higher energies.

On top of the EUI map in Fig.\,\ref{fig:diff-look-directions}, the STIX reconstructed image in the energy range from 4 to 8 keV is shown with black contours (50, 70, and 90\%). In order to have sufficient counts for the image reconstruction, we integrated in time for the entire flare duration, which resulted in a total of about 1100 counts above the non-solar background of about 2600 counts. For the reconstruction we only included the subcollimators 10 to 6 and we used the CLEAN algorithm \citep{1974A&AS...15..417H}, with a CLEAN beam width of 45 arcsec, which reflects the spatial resolution of subcollimators 6. In light blue, we indicate the error on the source location as deduced from the visibility forward fitting algorithm \citep[VIS\_FWDFIT, ][]{2022A&A...668A.145V}. A preliminary STIX aspect solution \citep{2020SoPh..295...90W} has been applied, but the currently available accuracy is only accurate within about 30 arcsec. After the reconstruction of the STIX image, we co-aligned the STIX source with the brightest pixel, which should approximately indicate the location of the heated flare loop by a manual shift of $30$ arcsec.


\section{Interchange reconnection model \label{sec:discussions-energing-flx} \label{sec:discussions}}

In the following, we discuss the interchange reconnection model in the framework of the observations reported in the previous section. The summary plot in Fig.~\ref{fig:cartoon} connects the cartoon in panel (\emph{g}) describing the scenario with the actual observations of panels (\emph{a})-(\emph{f}).

During the pre-flare phase, we observed enhanced emission at different wavelengths, which STIX showed to be hot at roughly the same temperature of the flare. This indicates the presence of heated plasma even before the main nonthermal peak at 18:04:07 UT. 
This heated plasma is likely not due to chromospheric evaporation, since in panels (\emph{d}) and (\emph{e}), it is clearly seen above the limb, unambiguously locating the source above the flare loop.

After the STIX thermal peak, it is possible to observe the flare loop-top (panel (\emph{f})) as well as the above-the-loop region of emitting plasma, as shown in panels (\emph{b}) and (\emph{e}). This above-the-loop region is in good agreement with the base 
of the coronal jet. 
Moreover, this region has moved with respect to its initial location, during the pre-flare heating phase in panels (\emph{a}) and (\emph{d}), towards higher altitudes by about $2600\,\mathrm{km}$, which results in an average projected speed of about $30\,\mathrm{km}\,\mathrm{s}^{-1}$.
In the case of the interchange reconnection model, the emerging closed field lines come into contact with open field lines first at a given altitude and afterwards, in the course of time, the reconnection occurs with field lines higher up in the atmosphere \citep[e.g.,][]{1977ApJ...216..123H}.
Therefore, the emission that we observe above the flare loop may be the region where the electrons are accelerated, namely, at the reconnection site, as shown in panel  (\emph{e}). In such a case, the derived speed of the moving source is comparable to the values reported by previous studies \citep[e.g.,][]{2003ApJ...595L.103K} for the coronal source detected in HXRs. However, for this event, nonthermal HXR imaging was not possible due to limited counting statistics. 
For the sake of completeness, in panel (\emph{c}), the H$\alpha$ map shows chromospheric plasma being ejected, which may be interpreted as cool jet or surge. This is also consistent with the interchange reconnection model.


\section{Summary and conclusions \label{sec:conclusions}}

In this paper, we study the microflare SOL2021-02-18T18:05 of A8 GOES class associated with a coronal jet. This compelling event highlights the importance of the interchange reconnection model in the release of accelerated particles into interplanetary space. It is characterized by different key aspects that we summarize in the following:

\begin{itemize}

    \item A spectroscopic analysis by Solar Orbiter/STIX shows that this microflare has a prominent nonthermal component, considering that is an A GOES class event. The spectral index of the nonthermal fit with a power law index of $\delta = 3.9 \pm 0.3$ is rather hard compared to what is usually observed in microflares \citep[e.g.,][]{2008ApJ...677..704H}. However, such intriguing  microflares have been report before (\cite{2008A&A...481L..45H} and \cite{Ishikawa_2013}). Nevertheless, this event is an extreme case where particle acceleration works exceptionally well -- not a typical case. 

    \item The hard X-ray flare is temporally associated with a radio type III burst, which indicates that some of the accelerated electrons escape upwards from the acceleration site. As the type III burst is seen down to $\sim$1 MHz, it is clear that there are open field lines that connect the flare site into interplanetary space. 

    \item The nonthermal emission is also clearly correlated in time with the escape of a jet seen in EUV. The tie-pointing method has been used to track the trajectory of the jet revealing a curved trajectory away from the Sun. The jet could be followed up to an altitude of $0.15$ \(\textup{R}_\odot\) above the solar surface, at which point the jet is still moving upward and away from the Sun. As the interplanetary type III burst indicates that energetic electrons are escaping the Sun, it is most likely the case that also the jet is escaping. 

    \item The most important  new finding that complements the interchange reconnection picture is the discovery of a hot ($\sim$12 MK), compact source in EUV that is seen above the main flare loop. The source already appears in pre-flare phase, and it is also seen in X-rays. As the source is above the flare loop, it cannot be produced by heated evaporated plasma, but it is, rather, the result of heating in the corona associated with the reconnection process. The source moves toward higher altitudes in time with a velocity of $\sim 30\,\mathrm{km}\,\mathrm{s}^{-1}$, reflecting successive interchange reconnection at higher and higher altitudes. Furthermore, the source corresponds to the starting point of the jet. All these observational findings indicate that the EUV source outlines the energy release region around the reconnection site within the interchange reconnection model (see cartoon in Fig.~\ref{fig:cartoon}(\emph{g})).  The detection of this source is made possible as AIA sees the flare at the limb and the hot source is clearly seen above the limb, unambiguously locating the source above the flare loop. An on-disk view of this flare, would have made  identification of this source very difficult due to projection effects, if not impossible. 

\end{itemize}

This single event study using multi-vantage point observations further corroborate that interchange reconnection indeed is a viable candidate for particle acceleration in the low corona on field lines open to interplanetary space. The event under discussion here, however, is an extreme case where electron acceleration works very efficiently and it might not be representative for events with less efficient magnetic energy conversion into nonthermal particles. A statistical analysis of jets and hard X-ray is needed to answer the question of whether such extreme events should be considered as proxy for less intense events happening all the time and everywhere on the Sun and for events with lower efficiency in the acceleration process. As the nonthermal counts are already low for this prominent flare, such a study would likely suffer from a sensitivity issue in hard X-rays. In any case, the best approach for such a study is to use STIX data taken a few days around Solar Orbiter perihelion. For future individual event studies, the next step is to find a similar event for which the escaping field lines intersect with Solar Orbiter or Parker Solar Probe. For such an event, the derived flare-accelerated electron spectrum can be compared with in situ measured spectra to further determine whether interchange reconnection is indeed a main contributor to the escape of energetic electrons from the Sun. 

\begin{acknowledgements}
      We wish to acknowledge the anonymous referee of this paper for the valuable inputs and encouraging comments. Solar Orbiter is a space mission of international collaboration between ESA and NASA, operated by ESA. The STIX instrument is an international collaboration between Switzerland, Poland, France, Czech Republic, Germany, Austria, Ireland, and Italy. The EUI instrument was built by CSL, IAS, MPS, MSSL/UCL, PMOD/WRC, ROB, LCF/IO with funding from the Belgian Federal Science Policy Office (BELSPO/PRODEX PEA 4000112292); the Centre National d’Etudes Spatiales (CNES); the UK Space Agency (UKSA); the Bundesministerium für Wirtschaft und Energie (BMWi) through the Deutsches Zentrum für Luft- und Raumfahrt (DLR); and the Swiss Space Office (SSO). We acknowledge the use of data from the Solar X-ray Monitor (XSM) on board the Chandrayaan-2 mission of the Indian Space Research Organisation (ISRO), archived at the Indian Space Science Data Centre (ISSDC). XSM was developed by Physical Research Laboratory (PRL) with support from various ISRO centers. We thank Fachhochschule Nordwestschweiz (FHNW), Institute for Data Science in Windisch, Switzerland, for hosting the e-Callisto network.
      
      AFB, HC and SK are supported by the Swiss National Science Foundation Grant 200021L\_189180 for STIX. JS and AMV acknowledge the Austrian Science Fund (FWF): I4555-N. The work of FS was supported by DLR grant No. 50 OT 1904. 
      Thanks are also extended for valuable input by Sophie Musset and Slimane Mzerguat.
\end{acknowledgements}

%
%

\bibliographystyle{aa} 
\bibliography{aanda} 

\end{document}